# Straight contact lines on a soft, incompressible solid.

Laurent Limat

*Laboratoire Matière et Systèmes Complexes (MSC), UMR 7057 of CNRS and Paris Diderot University, 10 rue Alice Domon et Léonie Duquet, 75013 Paris, France.*

**Abstract:** The deformation of a soft substrate by a straight contact line is calculated, and the result applied to a static rivulet between two parallel contact lines. The substrate is supposed to be incompressible (Stokes like description of elasticity), and having a non-zero surface tension, that eventually differs depending on whether its surface is dry or wet. For a single straight line separating two domains with the same substrate surface tension, the ridge profile is shown to be be very close to that of Shanahan and de Gennes, but shift from the contact line of a distance equal to the elastocapillary length built upon substrate surface tension and shear modulus. As a result, the divergence near contact line disappears and is replaced by a balance of surface tensions at the contact line (Neumann equilibrium), though the profile remains nearly logarithmic. In the rivulet case, using the previous solution as a Green function allows one to calculate analytically the geometry of the distorted substrate, and in particular its slope on each side (wet and dry) of the contact lines. These two slopes are shown to be nearly proportional to the inverse of substrate surface tensions, though the respective weight of each side (wet and dry) in the final expressions is difficult to establish because of the linear nature of standard elasticity. A simple argument combining Neumann and Young equations is however provided to overcome this limitation. The result may have surprising implications for the modelling of hysteresis on systems having both plastic and elastic properties, as initiated long ago by Extrand and Kumagai.

# 1 – Introduction.

Wetting of a deformable solid has been investigated long ago by many authors [1-9], and is presently attracting again a great attention [10-24]. Experimental studies are still very active on different soft substrates (elastomers, gels, polymers…) [10-20] in a field having potentially a large number of applications (soft lenses, biomaterials, low friction surfaces…), while new insights have been provided recently in the modelling of contact lines on soft substrates [18, 20-23]. In the original vision, developed by de Gennes and Shanahan [3-5], it is known that a "ridge" is formed by the liquid surface tension $\gamma$ pulling the substrate near the contact line, whose dimensions scale as $\gamma/E$, E being the Young modulus of the substrate (see fig.1). More precisely, the dominant contribution of the surface distortion should read:

$$\zeta(x) \approx \frac{2(1-\nu^2)}{\pi} \frac{\gamma \sin\theta_0}{E} Log\frac{\Delta}{|x|} \quad , \quad (1)$$

in which x is the distance to the contact line, $\zeta(x)$ the normal displacement of the substrate surface, $E$ the Young modulus of the substrate, $\nu$ the Poisson ratio, $\Delta$ a macroscopic cut-off and $\theta_0$ the contact angle at which the liquid surface meets the undistorted solid. The ridge should thus have a logarithmic profile that diverges for small and large x, divergences that were supposed to be screened at large x by some macroscopic scale $\Delta$ (typically the drop size), and at small x by plastic effects (or non-linearities). In turn, the interaction of the ridge profile with these plastic effects were also invoked to explain the increase of dissipation of a contact line moving on soft solids [5-7,9], and also a part of the wetting hysteresis exhibited by soft compounds [5,8].

More refined approaches have emerged recently [18, 20-23] taking into account the substrate surface tension, that, except in a variational description [4] was usually forgotten. The profile (1) is modified [18,20], the elastocapillary length [25] $l_S \sim \gamma_S/E$ built upon surface tension of the substrate and its elastic modulus entering into the problem. In addition, the divergence of (1) is replaced by a Neumann condition of balance between surface tensions at the contact line analogous to the well known case of liquid/liquid systems [18]. This liquid/liquid behaviour is expected to hold at scales smaller than $l_S$, while at larger scale a cross-over towards standard Young wetting of solids is supposed to take place [22]. Non-trivial effects are also expected in the redistribution of capillary forces from the liquid to the solid [20,21]. A general problem in all these references is that the calculations are very complex when one combines the full theory of elasticity with the axisymmetric geometry of sessile drops, and sometimes with inputs coming from microscopic modelling. For instance, this complexity, except in qualitative reasoning on energies used to predict transitions between "solid like" and "liquid like" behaviours of the substrate [22], has constrained most of recent quantitative elastowetting calculations to treat only the case of a macroscopic contact angle equal to 90°, i.e. to consider a solid having the same surface tension, in the dry part of its surface $\gamma_S$, as well as in the wetted one $\gamma_{SL}=\gamma_S$.

In the present paper I suggest an alternative approach that allows one to remain very close to the initial idea of Shanahan and de Gennes, keeping its relative simplicity, but this simplicity allows in turn, step by step, to address situations as the one imagined here in better conditions (rivulet with a substrate surface tension mismatch). First, rather than using the full theory of elasticity, I focus on the incompressible case *(ν=1/2)*, which corresponds in fact to most of soft solids of interest (gels, elastomers...) and use a Stokes like description of elasticity, in which the pressure is treated specifically [26]. This will allow us to use interesting analogies [27] between elastic fields and Stokes flows [28]. Also, rather than targeting situations with a curved contact lines (as occurs for a sessile 3D drop), I rather focus on the case of straight contact lines, here on a substrate of infinite depth.

Using two different methods (one qualitative and one rigorous), I first show that the exact solution for a infinite straight contact line on a substrate of surface tension $\gamma_S$ is very close to a generalisation of (1) in which the dependence on x is simply shift by a constant proportional to the elastocapillary length. This removes the divergence of the solution at small scale, that is replaced by a balance of surface tensions on contact line, as found by other authors [18,22,23]. I then use this solution as a "Green function" that allows one to calculate analytically the deformations induced now by two parallel contact lines connected by a curved liquid free surface, i.e. a rivulet of infinite length but of finite lateral extent *R* that can be related explicitly to the macroscopic cut-off. Finally, treating the variation of surface tension between the "dry" and "wet" values as an additional film covering the wetted domain of surface tension $\delta\gamma=\gamma_{SL}-\gamma_S$, I also address the rivulet problem without restriction on the wettability of the substrate. Though not applicable directly to sessile drops with a circular perimeter, this method allows for physical discussions and to a first exploration of a situation with a contact angle having non simple values.

The perhaps most important result that we will get here is the slope of the substrate near the contact line (see eqs. 38 below) that we explicitly relate to the substrate surface tension. This result is important for wetting studies on systems having both elastic and plastic properties [8,13,14,24], where the slope taken by the substrate is an important parameter to predict onset of depining of contact lines. These aspects are discussed in the last section of the paper, where more speculative views on hysteresis of contact angle are also considered. The results presented here are not only interesting for these specific questions. Solving the case of straight infinite contact lines is important for several wetting geometries, and for instance experiments involving dip coating [14,29] . This situation is one of the sole allowing careful studies of dynamical effects without the complexity introduced by perimeter shape selection on expanding or moving drops [24,30].

**2 - Outline of the paper.**

In the present work, I thus reconsider the problem of a soft – but incompressible - elastic solid distorted by the surface tension of a liquid deposited at its free surface, taking into account the substrate surface tension. For simplicity, I will consider only a 2D geometry (x,y), invariant in the z-direction that will correspond to a single straight contact line and later to a liquid ridge or rivulet deposited on the solid, combining two parallel contact lines. To keep small in the calculations the local slope of the substrate surface (which is required by standard theory of elasticity) I will also assume that the liquid surface tension $\gamma$ remains small compared to the substrate surface tension. This is a strong limit of our modelling, but will be enough to understand what happens at least qualitatively in more general situations.

In section 3, I first investigate the case of a single contact line, in the case of a "symmetrical" surface tension distribution of the substrate, i.e. $\gamma_{SL}=\gamma_S$ where $\gamma_S$ and $\gamma_{SL}$ designate respectively the surface tension of the dry and wet substrate. I show that a logarithmic solution emerges again but shift, and then screened at small scale by the elasto-capillary length $l_S=(1/\pi)(\gamma_S/\mu)$. At this scale, the divergence of stresses invoked by Shanahan and de Gennes is replaced by a Neumann condition of equilibrium between the liquid and solid surface tensions, just as what occurs for liquids. This result is reached in subsection 3-1 via a simplified argument in terms of scaling laws, while in subsection 3-2 a full rigorous calculation is presented, in which the effect of surface tension is treated as a Dirac function, with a decomposition over Fourier modes. In subsection 3-3, I also investigate the structure of the horizontal displacements of the substrate. These ones vanish when the contact angle is equal to the equilibrium "thermodynamic" value imposed by Young principle as well as by symmetries $\theta_Y=\pi/2$. Forcing a different values for the contact angle is however mathematically possible but leads to a singularity of the displacement field near contact line, similar to the one imagined by Shanahan and de Gennes for the vertical displacements in the absence of substrate surface tension, with the classical choice between a infinite distortion energy or the occurrence of yield and plasticity near contact line.

In section 4, I use the approximate logarithmic solution found in the previous section to solve the case of a 2D drop (a rivulet) composed of two parallel contact lines connected by a cylindrical liquid free surface. Cumulating the effect of the two contact lines and of the Laplace pressure below the drop, I find the shape of the solid substrate, which gives analytically the height of the ridge, the deepness of the substrate depletion formed below the drop by Laplace pressure enclosed inside the 2D drop, the slope of the substrate near the contact lines, etc... Interestingly, the macroscopic cut-off introduced in the previous section disappears and is replaced by the distance separating the two contact lines, which attests of the consistency of the approach.

In the first subsection 4-1, I address the case of a symmetrical solid surface tension distribution, i.e. again the case $\gamma_S=\gamma_{SL}$ in which the surface tension of the wetted solid is equal to the dry value, the more complex case of a asymmetrical distribution being explored in subsection 4-2. In this part, to account for the change of surface tension between wet and dry parts of the surface $\delta\gamma=\gamma_{SL}-\gamma$, I introduced a "skin" of effective surface tension $\delta\gamma=\gamma_{SL}-\gamma_S$ deposited on the wetted part of the substrate, this "skin" adding an extra Laplace pressure that needs to be integrated over the whole wetted area of the substrate. The situation remains cumbersome, as one is faced with non-local equations ruling the substrate surface distortion, but several analytical results can be obtained in the limits $|\delta\gamma|<<\gamma_{SL}\sim\gamma_S$ and $R>>l_S$ , concerning the height of the ridge and the deepness of the Laplace pressure induced depletion below the drop. In addition, results will be also obtained concerning the limit slope of the substrate near the dry and wet sides of the contact line that are nearly equal to one half of the ratio between the vertical tension applied there by the liquid and the value of the relevant substrate surface tension (see eqs 38. below).

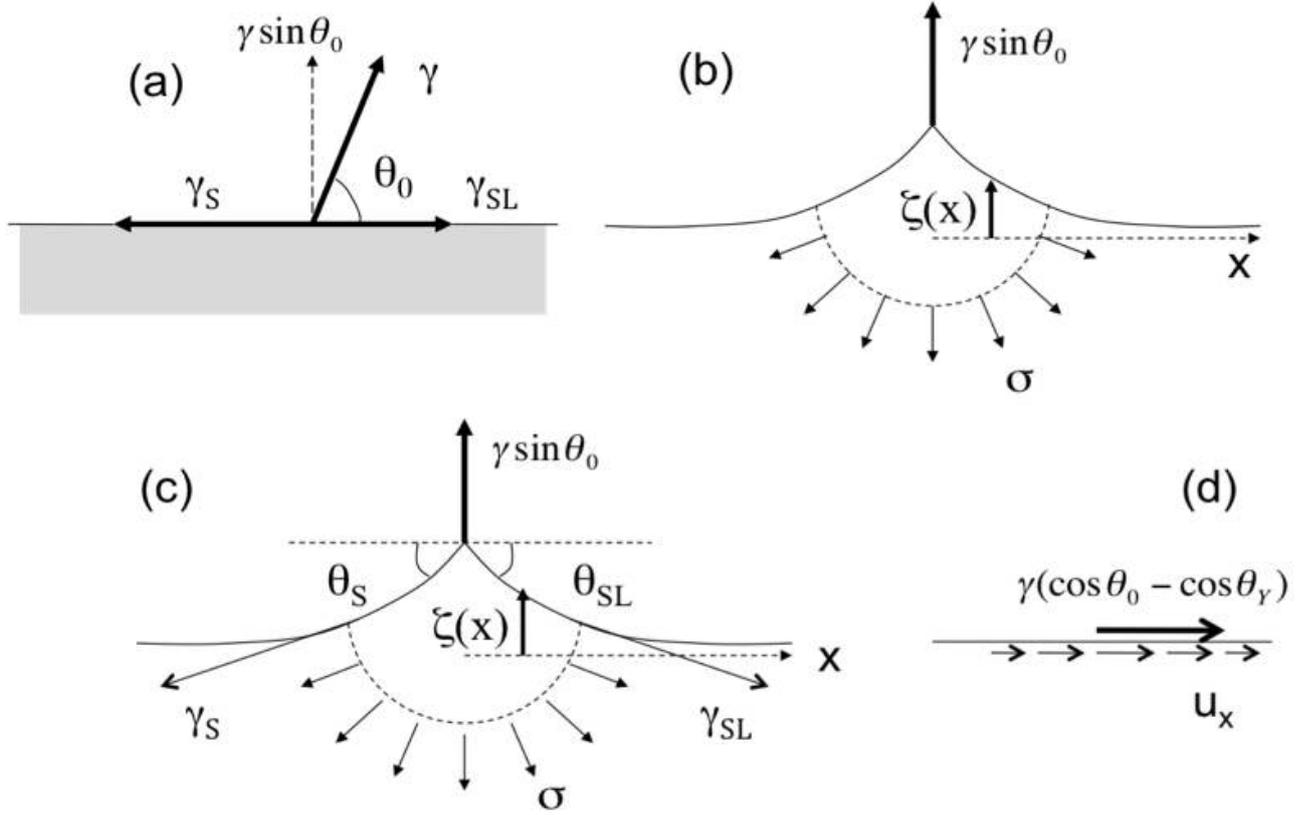

*Fig. 1*: (a) In the standard description of elastowetting, the surface tension are supposed to balance in a direction parallel to the substrate, the angle $\theta_0$ being equal to the Young value $\theta_Y$. On the other hand, (b) the normal component of liquid surface tension $\gamma\sin\theta_0$ is balanced by elastic stresses developing in the material. (c) For very soft materials, the surface distortion will add a vertical component of the substrate surface tension that will modify the previous equilibrium. (d) When $\theta_0$ differs from $\theta_Y$ the tangent mismatch of surface tensions develop an additional singular elastic field but the surface motions remain nearly parallel to the surface for a incompressible substrate.

From a physical point of view, each side of the ridge collecting one half of the force exerted by the liquid interface, force that is *in fine* communicated to the solid across the Laplace pressure drop associated to the substrate surface tension. In section 4-3, I compare the result with an argument inspired by the recent paper from Style and Dufresne [22], assuming that Neumann condition should hold also in the horizontal direction at contact line, while the Young angle should be recovered from a large scale point of view. The result obtained is very close, but slightly different, as we are handling two small parameters here, the ratios $\gamma/\gamma_S$ and the relative mismatch $\delta\gamma/\gamma_S$, the corrective term being of order of the product of both parameters. I show that this limitation is linked to the linear nature of standard elasticity, the exact result requiring to use a complex finite distortion method, out of the scope of the present paper. However, using the ideas of refs[22], I propose a simple argument allowing one to recover the exact result, which is not so far from our initial finding.

In section 5, I reconsider a mechanism proposed years ago by Extrand and Kumagai [8] to interpret at least a part of hysteresis on soft substrates. In short, the idea is to admit that in certain conditions (presumably the viscoelastic rheology is important), all seems to happen as if the liquid had to depin from the ridge he has created, the depining condition being simply an advancing or receding condition on a inclined surface (i.e. the distorted surface of the ridge). I first remind the initial argument in subsection 5-1, and then show how our findings would imply to modify it in subsection 5-2. A puzzling result that emerges at the very end is that, though this hysteresis has its origin in elastic and plastic effects, its final value could in fact depend only on the ratios of surface tensions.

The results of section 3 are consistent with those obtained recently by Jerison et al [18], and Das et al. [20] with respect to the emergence of the elasto-capillary length at small scale. In comparison with these works, the simplifications that we have introduced (incompressibility, infinite depth, 2D planar geometry) has allowed us to address the whole problem only analytically. Note also that we did not follow the suggestion of Das et al [20], about the possible existence of an extra-force perpendicular to the contact line, having its origin in the long range microscopic forces. The simple modelling that we have here in mind is to consider the solid as simply as possible, i.e. as an elastic solid, whose "skin" can be identified to a isotropic membrane under tension, this tension being possibly different for the dry and wet part of the substrate. In particular, we do not distinguish between the notions of "surface stress" and "surface energy" that one can find in ref [31,32], the isotropy approximate being supposed to hold for systems in some sense close to the liquid state.

## 3 – Single contact line with symmetry of the substrate surface tensions.

The geometry of the problem is suggested on Fig. 1. We investigate the response of the substrate to a localized force $\gamma sin\theta_0$ pulling it from above in the "vertical" direction (perpendicular to the substrate surface), the three surface tensions balancing in the direction parallel to the surfaces, when the Young condition is satisfied, i.e. $\theta_0=\theta_Y$, with:

$$\gamma_S = \gamma_{SL} + \gamma \cos\theta_Y \qquad (2)$$

Eventually, we will also discuss the situation when this condition is not satisfied, which implies, as suggested on Fig. 1-d the appearance of horizontal displacements of singular nature near contact line, induced by the horizontal component of surface tensions $\gamma(\cos\theta_0 - \cos\theta_Y)$ (see later, section II-C). For simplicity, we treat in the present section II only the case in which the two substrate surface tensions are equal ( $\gamma_{SL}=\gamma_S$ ). We first develop a qualitative argument and then a more rigorous calculation.

Note that, to calculate the effect of $\gamma \sin\theta_0$ when $\gamma_{SL}=\gamma_S$, we can assume a symmetrical distribution of stress induced by this vertical force, even when $\theta_0$ is not equal to $\pi/2$. As we shall see in sectionS 3-2 and 3-3, the fields induced by each component $\gamma \sin\theta_0$ and $\gamma \cos\theta_0$ decouple in the limit of a incompressible medium, the second one introducing only "horizontal" displacements at the free surface.

### 3-1. Qualitative argument.

In the limit of a incompressible solid, Shanahan expression (1) reduces simply to:

$$\zeta(x) = \frac{\gamma \sin\theta_0}{2\pi\mu} Log \frac{\Delta}{|x|} \tag{3}$$

where $\mu=E/[2(1+\nu)]=E/3$ is the shear modulus. This result can be recovered qualitatively by assuming that the vertical traction $\gamma \sin\theta_0$ develops in the bulk of the material a typical stress $\sigma$, at the surface of the cylinder of radius x suggested on Fig. 1-b given by the balance of vertical forces $\gamma \sin\theta_0 \propto x\sigma$. In fact this stress depends on both x and the angular coordinate, but at the free surface should reduce to the shear term $-\mu \partial \zeta/\partial x$, which yields the approximate equation

$$\frac{\partial \zeta}{\partial x} \approx -\alpha \frac{\gamma \sin\theta_0}{\mu x} \tag{4}$$

that can immediately be integrated to recover (3), which gives us the value of the unknown constant $\alpha=1/(2\pi)$. Let us now try to include in the description, the surface tension of the substrate. Standard elasticity being a first order linear theory, we consider here only slight deflections of the substrate free surface, i.e. $\theta_s(x) \approx \partial \zeta/\partial x << 1$. Considering fig. 1-b, it is obvious that the surface tension of the substrate should have been included in the vertical balance of forces which reads in fact for x>0:

$$\gamma \sin\theta_0 = -2\gamma_S \frac{\partial \zeta}{\partial x} - 2\pi \mu x \frac{\partial \zeta}{\partial x} \tag{5}$$

which finally yields the following expressions for $\zeta(x)$, and for the local slope of the substrate $\theta_s(x) \approx \partial \zeta/\partial x$:

$$\zeta(x) = \frac{\gamma \sin\theta_0}{2\pi\mu} Log \frac{\Delta + l_S}{|x| + l_s} \tag{6-a}$$

$$\theta_s(x) \approx \zeta'(x) = \frac{\gamma \sin\theta_0}{2\pi\mu} \frac{1}{l_S + |x|} \tag{6-b}$$

where $l_S=(1/\pi)(\gamma_S/\mu)$ designates an elastocapillary length built upon surface tension and elastic modulus of the substrate. Note that, though this expression is still singular for x=0, this singularity is considerably reduced. $\zeta(x)$ does *not* diverge for *x=0* and reaches a finite value. Also, the slope of the substrate on each side of the contact line is finite and given by $\pm\theta_S$ with:

$$\theta_S = \frac{\gamma \sin\theta_0}{2\pi\mu l_S} = \frac{\gamma}{2\gamma_S}\sin\theta_0 \tag{7}$$

where, of course, $\theta_S$ is supposed to be small, which implies that our modelling is correct only in the limit $\gamma \ll \gamma_S$, as announced in section 2. Anyway, we observe that Neumann equilibrium condition of surface tensions is satisfied in the vertical direction as $\gamma \sin\theta_0 \approx 2\gamma_S\theta_S$. If we can summarize the physics at play here, all seems to happen as if the surface tension of the liquid was redistributed at the contact line over each of the two substrate surfaces, i.e. on each of the surface tensions of the wetted and dry zones of the substrate, the resulting forces being in turn balanced by the elastic stresses developing in the bulk via the Laplace pressure difference formed between each side of each interface.

**3-2. Exact calculation.**

In the standard, general, theory of elasticity, the displacement fields $u_i(x_j)$ obeys the following equation:

$$\rho \frac{\partial^2 u_i}{\partial t^2} = \frac{\partial \sigma_{ij}}{\partial x_j} \tag{8}$$

with $\sigma_{ij} = \lambda \varepsilon_{kk}\delta_{ij} + 2\mu\varepsilon_{ij}$ and $\varepsilon_{ij} = (1/2)(\partial u_i/\partial x_j + \partial u_j/\partial x_i)$, $\lambda$ and $\mu$ being the Lamé coefficients of the substrate and $\rho$ the mass density. Eliminating the stress tensor in these equations, one usually gets an equation in terms of only the displacements that reads:

$$\rho \frac{\partial^2 \vec{u}}{\partial t^2} = \mu \Delta \vec{u} + (\lambda + \mu)\vec{\nabla}(\vec{\nabla}\cdot\vec{u}) = 0 \tag{9}$$

in the static case. Treating the incompressibility limit with this equation is puzzling as one should have simultaneously $\vec{\nabla}\cdot\vec{u} = 0$ and $\lambda = \nu\mu/(1/2 - \nu)$ infinite… A simple idea here consists in reducing this equation to its Stokes equivalent in fluid mechanics, after treating the second term as a pressure gradient, the effective pressure $P = -(1/3)\sigma_{kk} = -(\lambda + 2\mu)(\vec{\nabla}\cdot\vec{u})$ being removed from the stress tensor, in turn only ruled by the shear modulus. This incompressibility limit of elasticity theory is scarcely used, despite some allusions in Love[26], and recent works on viscous folding [27]. In analogy with Stokes equations of fluid mechanics [28], one has just to solve the two coupled equations:

$$\begin{cases} \nabla \cdot u = 0 \\ 0 = \mu\Delta u - \nabla P \end{cases} \tag{10}$$

completed at the boundary by the continuity of the full stress tensor that reads:

$$\sigma_{ij} = -P\delta_{ij} + 2\mu\varepsilon_{ij} \tag{11}$$

To build something more precise than a qualitative argument, one can solve now exactly the problem, by considering a surface tension force, of components ($\gamma cos\theta_0$, $\gamma sin\theta_0$) applied to an elastic, incompressible, substrate at a point of coordinates *(x=0, y=0)*, the substrate having a non-zero surface tension $\gamma_S$. In the general case, the boundary conditions should read at the surface of the solid substrate:

$$\begin{cases} \sigma_{yy} = \gamma\sin\theta_0\delta(x) + \dfrac{d}{dx}\left[\gamma_S\dfrac{d\varsigma}{dx}\right] \\ \sigma_{xy} = \gamma\cos\theta_0\delta(x) + \dfrac{d}{dx}[\gamma_S] \end{cases} \tag{12}$$

in which $\delta(x)$ is the Dirac function. Here, $\gamma_S$ is uniform and has the same value on both side of the contact line, which simplifies greatly the two conditions that read finally:

$$\begin{cases} \sigma_{yy} = -P + 2\mu\dfrac{\partial u_y}{\partial y} = \gamma\sin\theta_0\delta(x) + \gamma_S\dfrac{d^2\varsigma}{dx^2} \\ \sigma_{xy} = \mu\left(\dfrac{\partial u_x}{\partial y} + \dfrac{\partial u_y}{\partial x}\right) = \gamma\cos\theta_0\delta(x) \end{cases} \tag{13}$$

These boundary conditions must be completed by bulk equations that are the incompressibility and the equilibrium equations (9) above. To satisfy automatically the zero-divergence equation, we introduce here an "effective" stream function defined as:

$$u_x = -\frac{\partial\psi}{\partial y} \qquad u_y = \frac{\partial\psi}{\partial x} \tag{14}$$

The strategy of the calculation will then consist in developing the Dirac function above, as well as the other quantities over Fourier modes upon x, the bilaplacian nature of $\psi(x,y)$, and the laplacian nature of the pressure field allowing us to solve very easily the dependence upon y. In short:

$$\begin{aligned} \delta(x) &= \frac{1}{2\pi}\int dk\, e^{ikx} \\ \varsigma(x) &= \frac{1}{2\pi}\int dk\, \varsigma_k e^{ikx} \\ P(x,y) &= \frac{1}{2\pi}\int dk\, P_k e^{ikx} e^{|k|y} \\ \psi(x,y) &= \frac{1}{2\pi}\int dk\, (A_k + B_k y) e^{ikx} e^{|k|y} \end{aligned} \tag{15}$$

The complex amplitudes $\varsigma_k$ and $A_k$ are linked by the continuity of vertical displacements at the free surface of the solid, which imposes $\varsigma_k = ikA_k$, while the balance of pressure and elastic shear stress in the bulk (eq. 10-b) yields $P_k = 2i\mu k B_k$. The continuity of transverse stresses at the solid surface leads now to

$$k^2 A_k + |k| B_k = -\frac{\gamma}{2\mu}\cos\theta_0, \qquad (16)$$

and the balance of normal stresses $2i\mu|k|kA_k = \gamma\sin\theta_0 - \gamma_S k^2$ leads to

$$\varsigma_k = \frac{\gamma\sin\theta_0}{2\mu|k| + \gamma_S k^2}, \qquad (17)$$

which finally yields the following integral expression for ζ(x):

$$\varsigma(x) = \frac{1}{2\pi}\frac{\gamma\sin\theta_0}{\mu}\int_{1/\Delta'}^{+\infty}\frac{\cos kx}{k + \frac{\gamma_S}{2\mu}k^2}dk \qquad (18)$$

where Δ' is again some cut-off introduced at large scales, to limit the divergence of the integral near k=0. Note that, in the very specific case of an incompressible substrate, considered here, the influence of a possible horizontal force γ cos θ$_0$ vanishes for the vertical displacements of the surface, as suggested on Fig.(1-d). This is consistent with Shanahan's calculations, though obtained in the absence of substrate surface tension [5]. The integral reads now:

$$I_1 = \int_{1/\Delta'}^{+\infty}\frac{\cos kx}{k + \frac{\gamma_S}{2\mu}k^2}dk = -Ci\left(\frac{x}{\Delta'}\right) + \cos\left(\frac{2}{\pi}\frac{x}{l_S}\right)Ci\left(\frac{x}{\Delta'} + \frac{2}{\pi}\frac{x}{l_S}\right) - \frac{\pi}{2}\sin\left(\frac{2}{\pi}\frac{x}{l_S}\right) + \sin\left(\frac{2}{\pi}\frac{x}{l_S}\right)Si\left(\frac{x}{\Delta'} + \frac{2}{\pi}\frac{x}{l_S}\right) \qquad (19)$$

where Si and Ci designates respectively the Sine Integral and Cosine Integral functions defined as:

$$Si(x) = \int_0^x \frac{\sin t}{t}dt \qquad Ci(x) = -\int_x^{+\infty}\frac{\cos t}{t}dt \qquad (20)$$

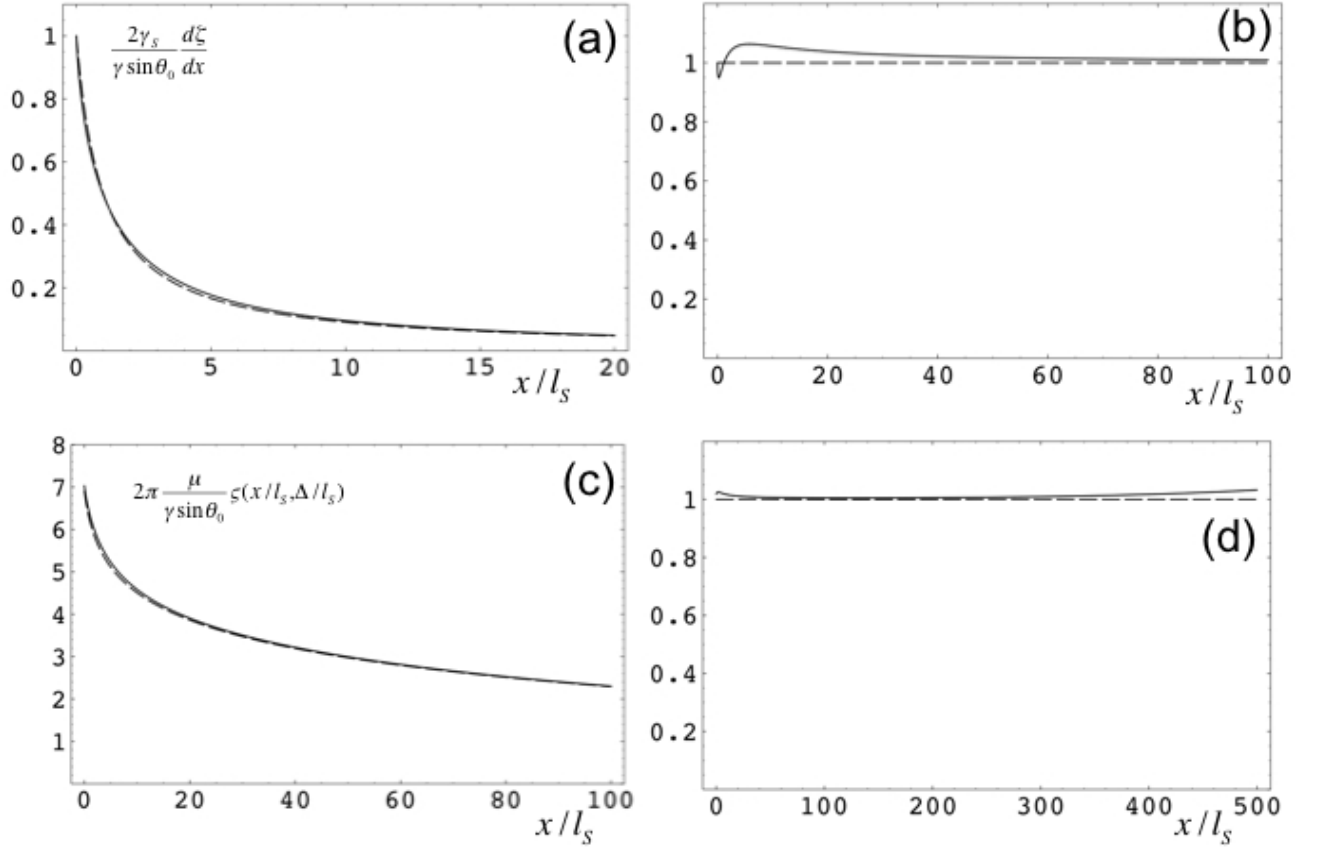

**Fig. 2**: (a) Comparison between the approximate slope distribution given by eq. 20 (dashed line) and the exact one of eq. 5 (continuous line, in the half space x>0. (b) Ratio of the exact slope distribution to the approximate one. (c) Comparison between the exact profile of the distorted surface with $\Delta/l_S=1000$ (continuous line), to the approximate one (dashed line) with $\Delta'/\Delta=1.7807$. (d) ratio of the exact surface profile to the approximate one.

From (20), one can also deduce easily the slope of the substrate as a function of x:

$$\frac{d\zeta}{dx} = \frac{\gamma \sin\theta_0}{2\gamma_S}\left[\frac{2}{\pi}Ci\left(\frac{2}{\pi}\frac{x}{l_S}\right)\sin\left(\frac{2}{\pi}\frac{x}{l_S}\right) + \cos\left(\frac{2}{\pi}\frac{x}{l_S}\right)\left(1 - \frac{2}{\pi}Si\left(\frac{2}{\pi}\frac{x}{l_S}\right)\right)\right] \quad (22)$$

It is interesting here to note that the large scale cut-off $\Delta'$ has disappeared from this expression, just as for what would occur for the approximate logarithmic solution. I have compared on Figs. 2-a and 2-b, this exact solution (22) to its equivalent of the previous qualitative part (eq. 6-b):

$$\frac{d\zeta}{dx} \approx \mp\frac{\gamma \sin\theta_0}{\gamma_S}\frac{l_S}{|x|+l_S} \quad (20)$$

As one can check, though the two solutions are not exactly identical, they are very close to each other within an accuracy of a few percents. It is also important to realize that both expressions (the approximate and the exact ones) have the same values for x=0, which means that both satisfy Neumann equilibrium of surface tensions in the vertical direction. The same comparison can be done for the function ζ(x) itself, after noticing that in the limit $l_S<<x<<\Delta$', one has $I_1 \approx \gamma_{EM} + Log(\Delta'/|x|)$ where $\gamma_{EM} \approx 0.5772$ is the "gamma" Euler-Mascheroni constant. This imposes the relationship $\Delta' = \Delta e^{\gamma_{EM}} \approx 1.7807\Delta$ for consistency. With this choice of constants, the comparison is again very good on figs (2-c) and (2-d), which is enough here to validate the approximate logarithmic profile (6) modified by the elasto-capillary length, that we will use most often in the following sections.

**3-3. Horizontal displacements and the selection of Young angle.**

If we consider now the displacements parallel to the unstrained substrate $u_x = -\partial\psi/\partial y$, these ones read:

$$u_x(x, y=0) = \frac{\gamma \cos\theta_0}{4\pi\mu} \int dk \frac{e^{ikx}}{|k|} \tag{23}$$

In principle, the integral is diverging at both large and small k. The small k limit can be regularized by the macroscopic scale available (here the width of the rivulet), but the large k limit will remain. For a perfectly elastic material, the sole possibility to cancel the large k divergence, that would introduce a infinite elastic energy, is to assume that the whole system (a whole drop for instance, deposited on a plane, as on Fig. 3, or the cross-section of a rivulet) will select $\theta_0=\pi/2=\theta_Y$, which means the Young, thermodynamic, equilibrium contact angle, that is the sole to allow an exact balance of surface tensions in the direction parallel to the substrate.

Now, it is interesting to remark that the problem has exactly the same structure as that found by Shanahan in the vertical direction when one neglects the substrate surface tensions, and we can here examine what would happen here with the same physical ideas. Most of soft materials are not only elastic but exhibit also plastic properties (or non-linear response), that could limit at large k the divergence of this integral. For simplicity, if we assume here a oversimplified rheology in which the elastic resistance to deformation vanishes when some maximal critical stress $\sigma_M$ is exceeded, (23) will reduce to:

$$u_x(x,y=0) = \frac{\gamma \cos\theta_0}{2\pi\mu} \int_{1/\Delta'}^{1/l_P} dk \frac{\cos kx}{k} = \frac{\gamma \cos\theta_0}{2\pi\mu} \left[ Ci(x/\Delta') - Ci(x/l_P) \right] \tag{24}$$

where $l_P$ is a cut-off that will be specified later. In the limit $l_P<<x<<\Delta'$, this expression reduces to an "à la Shanahan" limit that reads

$$u_x(x,y=0) \approx \frac{\gamma \cos\theta_0}{2\pi\mu} Log \frac{\Delta}{|x|} \tag{25}$$

with again $\Delta' = \Delta e^{\gamma_{EM}} \approx 1.7807\Delta$. We can now estimate the plastic scale $l_P$ by writing that when $x \approx l_P$ the typical stress $\mu\partial u_x/\partial x \sim \mu\partial u_x/\partial x \sim \sigma_M$ which yields:

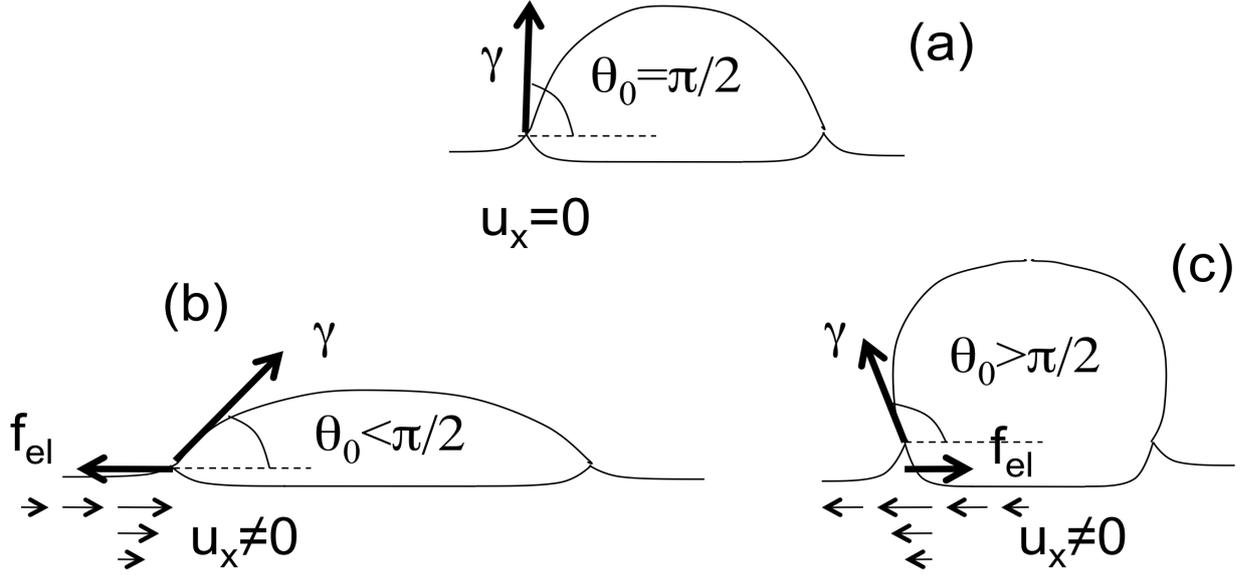

*Fig. 3: Depending on the history of a drop, and if the interaction between the contact line and the substrate allows some possible pinning effects, different states can occur even when the "natural" contact angle should be equal to π/2: (a) "ideal" state, with no displacements developing in a direction parallel to the substrate, (b-c) states "under tension" in which the capillary force is balanced by elastic forces.*

$$l_P \propto \frac{\gamma}{\sigma_M} |\cos\theta_0| \tag{26}$$

The situation to which we are faced with is suggested on figs. 3-a to 3-c. When a drop is lying on a soft substrate, with a "natural" contact angle close to $\pi/2$, depending on the history of the drop (impact, slow deposition, inflation or deflation via a syringe...), three different states can be in fact imagined.

In the "most natural" one, the drop adopts a hemispherical shape (in 3D, or cylindrical in 2D) with $\theta_0 = \pi/2$, while no horizontal displacement occur. Now, if one deflates or inflates the drop, the contact line can move, reproducing $\theta_0 = \pi/2$ with a different drop radius, but if the surface displays at small scale some rugosity or wettability defects, we can imagine that the contact line could also remained pinned, varying its apparent contact angle $\theta_0$. In these two new states, the drop exerts on the substrates horizontal forces of intensity $\gamma\cos\theta_0$ per unit length of contact line, that are balanced by forces exerted by the defects, and itself transformed into elastic stresses developing in the substrate, their divergence at small scale being eventually screened by plastic effects.

For a perfect elastic substrate, whose surface is also perfect (no rugosity, no wettability defects), the sole possible structure will be that of Fig. (1-a), i.e. the contact line will slide on the solid until $\theta_0 = \theta_Y = \pi/2$. If pinning is possible on the surface, and if the substrate can undergo a yield transition, the two other states could appear. Obviously, the same question can hold for any value of the Young contact angle. In what follows we will focus on the he simplest possibility, i.e. a situation in which the system converges towards the Young contact angle with no internal motion of the substrate in the horizontal direction.

## 4– Case of a liquid ridge (or rivulet) on the substrate.

We now consider the situation depicted on Fig. 4, of a ridge (or rivulet), infinite in the z-direction but of limited width 2R in the y direction, whose apparent contact angle coincides with the Young value ($\theta_0=\theta_Y$). There are several aspects that complicates the situation: presence of two contact lines instead of one, each pulling the substrate in the y-direction with a force per unit length equal to $\gamma sin\theta_0$, action on the region –R<x<R of a Laplace pressure $P_L = \gamma/a = (\gamma/R)sin\theta_0$ that tends to push the substrate downwards, and the possible discontinuity of the substrate surface tension at contact lines. Indeed, in the general case, the surface tension of the substrate/liquid interface $\gamma_{LS}$ is not the same as that of a air/substrate interface for which we will keep the symbol $\gamma_S$. The extension of the previous calculation to this case is drastically complex, and we will thus limit yourself to the simple limit $\gamma_{LS} = \gamma_S$ in the first subsection (4-1), exploring the case $\gamma_{LS} \neq \gamma_S$ in the next one (section 4-2). As $\theta_0=\theta_Y$, we will assume that surface tensions are balancing exactly at contact line in the horizontal direction without inducing an additional singular displacement field in the horizontal direction. We will thus only consider and superpose, the displacement fields induced by only the vertical components of stresses applied to the free surface of the substrate.

### 4-1 – Symmetric case, with $\gamma_{SL} = \gamma_S$.

In the spirit of eq. (6), that in some sense defines a "Green function" for the distortion of the substrate surface, the vertical displacement of this one should read $\zeta(x) = \zeta_S(x)$, where

$$\zeta_S(x) = \frac{1}{2\pi}\frac{\gamma}{\mu}\sin\theta_0 \left[ Log\frac{\Delta + l_S}{l_S + |x - R|} + Log\frac{\Delta + l_S}{l_S + |x + R|} - \frac{1}{R}\int_{-R}^{R} Log\frac{\Delta + l_S}{l_S + |x - x'|} dx' \right] \quad (29)$$

is the response of the substrate modified by the substrate surface tension $\gamma_S$ to the stresses exerted vertically. The two first terms are associated to each contact line where a normal tension is applied of intensity $\gamma sin\theta_0$, while the last term holds for the effect of the Laplace pressure enclosed inside the rivulet. After calculating the integral linked to Laplace pressure, this expression reduces to:

$$\zeta_S(x) = \frac{1}{2\pi}\frac{\gamma}{\mu}\sin\theta_0 \left[ \frac{l_S + x}{R} Log\frac{l_S + x + R}{l_S + x - R} - 2 \right] \quad \text{for x>R} \quad (30\text{-a})$$

and to:

$$\zeta_S(x) = \frac{1}{2\pi}\frac{\gamma}{\mu}\sin\theta_0 \left[ \frac{l_S}{R} Log\frac{l_S + x + R}{l_S} + \frac{l_S}{R} Log\frac{l_S + R - x}{l_S} + \frac{x}{R} Log\frac{l_S + x + R}{l_S + x - R} - 2 \right] \quad \text{for 0<x<R} \quad (30\text{-b})$$

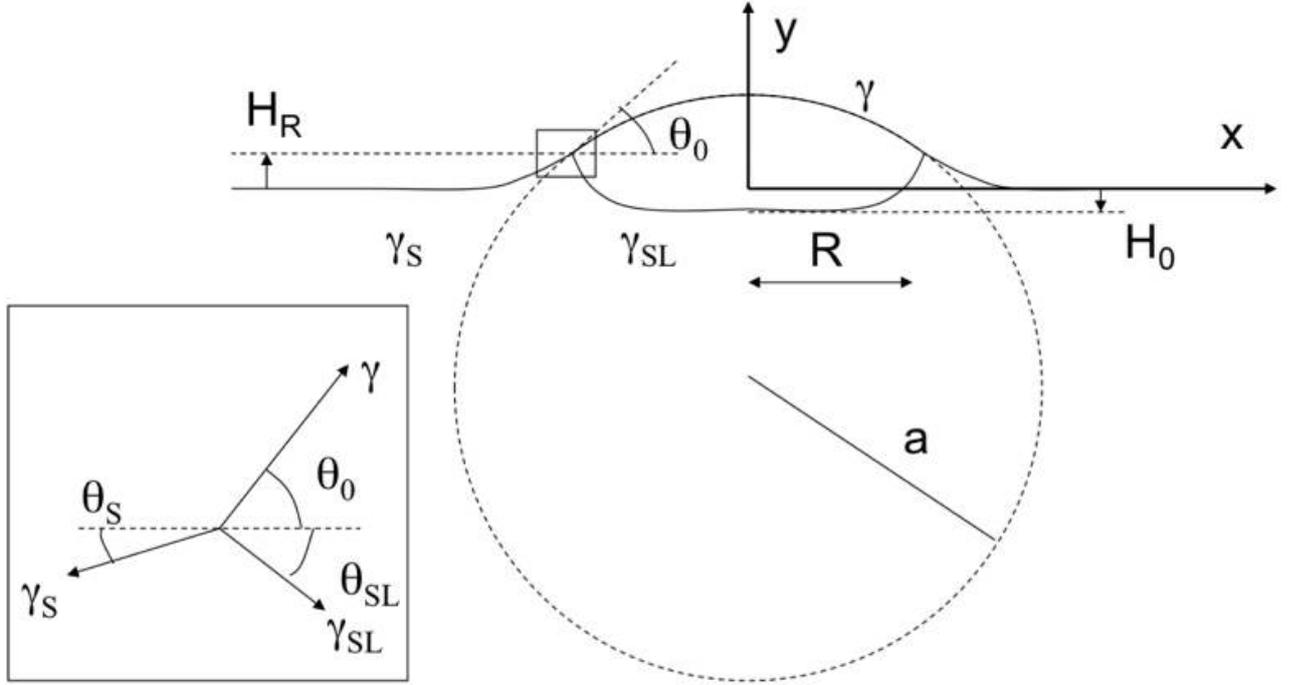

**Fig. 4:** *Notations and geometry involved in the problem of a liquid ridge (or rivulet) lying one a deformable substrate.*

As one can see, the large scale cut-off has disappeared, the new one being simply the width of the rivulet (!), while at small scales, the absence of divergences near both contact lines still holds because of the regularisation ensured by the elastocapillary length $l_S$. These expressions allow us to calculate both the height of the ridge $H_R$ and the vertical position of the drop center $H_0=\zeta_S(0)$. In the limit $R>>l_S$, these quantities reduce to:

$$H_R = \frac{1}{2\pi}\frac{\gamma \sin\theta_0}{\mu} Log\frac{R}{l_S} \qquad H_0 = -\frac{1}{\pi}\frac{\gamma \sin\theta_0}{\mu}[1 - \frac{l_S}{R} Log\frac{R}{l_S}] \qquad (31)$$

The structure of the first expression is rather well known, but the novelty here is that the large scale and small scale cut-off are explicitly related to respectively the "drop size" R and to the elastocapillary length. The second expression implies that, most often, the substrate is systematically lower inside the drop than outside because of the Laplace pressure developed below the free surface of the drop. It is also possible to calculate explicitly the slope of the substrate on each side of the contact lines. One finds respectively for the outer and inner slopes (both counted positively) as on fig. 3:

$$\begin{cases} \theta_S = \dfrac{\gamma \sin\theta_0}{2\gamma_S}\left[1+\dfrac{l_S}{2R+l_S}-\dfrac{l_S}{R}Log\left(1+\dfrac{2R}{l_S}\right)\right] \\ \theta_{LS} = \dfrac{\gamma \sin\theta_0}{2\gamma_S}\left[1-\dfrac{l_S}{2R+l_S}+\dfrac{l_S}{R}Log\left(1+\dfrac{2R}{l_S}\right)\right] \end{cases} \qquad (32)$$

As one can check, Neumann equilibrium in the vertical direction $\gamma \sin\theta_0 = \gamma_S \theta_S + \gamma_{SL}\theta_{SL}$ is automatically satisfied in this limit $\gamma_{SL}=\gamma_S$, despite a slight rotation of both the dry and wetted substrate induced by Laplace pressure in the drop, only visible for very small droplets. We are now going to try to escape a bit from this limit, by considering now the possibility a slight difference between the two surface tensions.

**4-2 – Asymmetrical case with $\gamma_{SL} \neq \gamma_S$.**

Including in the calculations two different substrate surface tensions in very difficult, especially for the Fourier method of section 4. A possibility to avoid this consists in using again our approximate Green function of equation (6-a), combined with an effective surface film, of surface tension $\delta\gamma=\gamma_{LS}-\gamma_S$ covering the whole wetted area, i.e. the region defined by $-R<x<R$. Note that we are not here playing with some uncontrolled fictitious object, but taking into account explicitly the change of surface tension below the liquid. After introducing our fictitious film, the vertical displacement should satisfy:

$$\zeta(x) = \dfrac{1}{2\pi}\dfrac{\gamma\sin\theta_0 - \delta\gamma\theta_{LS}}{\mu}\left[Log\dfrac{\Delta+l_S}{l_S+|x-R|}+Log\dfrac{\Delta+l_S}{l_S+|x+R|}\right]$$
$$-\dfrac{1}{2\pi}\dfrac{\gamma\sin\theta_0}{\mu}\dfrac{1}{R}\int_{-R}^{R}Log\dfrac{\Delta+l_S}{l_S+|x-x'|}dx' + \dfrac{1}{2\pi}\dfrac{\delta\gamma}{\mu}\int_{-R}^{R}\zeta''(x)Log\dfrac{\Delta+l_S}{l_S+|x-x'|}dx' \qquad (33)$$

that is just eq.(29) to which we have added the Laplace pressure exerted on the solid by the fictitious film (last term), and the vertical forces that are applied by this film on the two contact lines ($\delta\gamma\theta_{SL}$ contribution). After integrating by part the second integral, and eliminating the scale $\Delta$, one finally gets the correction $\delta\zeta=\zeta(x)-\zeta_S(x)$, that reads:

$$\delta\zeta(x) = \dfrac{1}{2\pi}\dfrac{\delta\gamma}{\mu}\int_{-R}^{R}\dfrac{d\zeta}{dx'}\dfrac{sgn(x'-x)}{l_S+|x'-x|}dx' \qquad (34)$$

where *sgn(x'-x)* designates the sign of *x'-x*. Note that (34) is not a first order development but a rigorous result, valid for any value of $\delta\gamma=\gamma_{SL}-\gamma_S$ (but of course while remaining in the limit $\gamma<<\gamma_S$ and $\gamma<<\gamma_{SL}$ of the whole paper). It seems difficult, for this first exploration of our method, to solve the obtained integral equation ruling $\zeta(x)$, but it is possible to calculate this extra term in the limit $\delta\gamma=\gamma_{SL}-\gamma_S << \gamma_S$, by using a perturbation method that consists in replacing $\zeta'(x')$ with $\zeta_S'(x')$ in the integral. The calculations are however so heavy that I have limited the effort to calculate the corrections on the quantities $H_R$ and $\theta_S$, for large values of $R/l_S$. I obtained:

$$\delta H_R \approx \frac{1}{2\pi} \frac{\delta\gamma}{\mu} \int_{-R}^{R} \frac{d\varsigma_S}{dx'} \frac{dx'}{l_S + R - x'} \approx \frac{1}{4\pi} \frac{\gamma \sin\theta_0}{\mu} \frac{\delta\gamma}{\gamma_S} \left[ 1 - \left(\frac{\pi^2}{3} + 1 - 2Log2\right)\frac{l_S}{2R} + \frac{l_S}{R} Log\frac{2R}{l_S} \right] \quad (35\text{-a})$$

$$\delta\theta_S \approx \frac{1}{2\pi} \frac{\delta\gamma}{\mu} \int_{-R}^{R} \frac{d\varsigma_S}{dx'} \frac{dx'}{(l_S + R - x')^2} \approx \frac{1}{4\pi} \frac{\gamma \sin\theta_0}{\mu R} \frac{\delta\gamma}{\gamma_S} \left[ Log\frac{2R}{l_S} - \frac{3}{2} \right] \quad (35\text{-b})$$

Combining the last expression with eqs.(30-a), one gets the complete expressions of the substrate slope near contact lines, on the dry substrate:

$$\theta_S \approx \frac{\gamma \sin\theta_0}{2\gamma_S} \left[ 1 + \frac{l_S}{2R} - \frac{l_S}{R} Log\left(\frac{2R}{l_S}\right) \right] + \frac{1}{4} \frac{\delta\gamma}{\gamma} \sin\theta_0 \frac{l_S}{R} \left[ Log\left(\frac{2R}{l_S}\right) - \frac{3}{2} \right] + \ldots \quad (36)$$

A surprise here is that in the limit $R \gg l_S$, i.e. for a "big" drop, the result simplifies and one gets simply the previous result $\theta_S = \gamma \sin\theta_0 /(2\gamma_S)$, in which any information on $\gamma_{SL}$ has completely disappeared, i.e. $\theta_S$ seems to depend only upon $\gamma_S$ and $\gamma$. Though this result is obtained here only at first order upon $\delta\gamma/\gamma$ it is in fact much more general. One can indeed deduce from eq. 34 the change in $\theta_S$ induced by the difference $\delta\gamma = \gamma_{SL} - \gamma_S$ that reads:

$$\delta\theta_S = \theta_S - \theta_S(\gamma_{SL} = \gamma_S) = -\frac{1}{2\pi} \frac{\gamma_{SL} - \gamma_S}{\mu} \int_{-R}^{R} \frac{\theta_{sl}(x')}{(l_S + x')^2} dx', \quad (37)$$

in which $\theta_{sl}(x) \approx \varsigma'(x)$ is the local slope of the substrate inside the drop. This expression goes to zero as $l_S/R$ when this ratio becomes asymptotically small. Combining this result, valid in the limit of very large drops, with the idea that, again in this limit, $\gamma_{LS}$ and $\gamma_S$ should play symmetrical role in the selection of $\theta_{LS}$ and $\theta_S$, one finally gets for these two quantities:

$$\begin{cases} \theta_S \approx \dfrac{\gamma \sin\theta_0}{2\gamma_S} \\ \theta_{SL} \approx \dfrac{\gamma \sin\theta_0}{2\gamma_{SL}} \end{cases} \quad (38)$$

Again, Neumann equilibrium $\gamma \sin\theta_0 = \gamma_S \theta_S + \gamma_{SL} \theta_{SL}$ is satisfied in the direction normal to the unperturbed substrate, and it is interesting to note here the non-trivial result that each side of the substrate surface (the dry one and the wetted one) supports exactly one half of the total normal force. Of course, as mentioned several times above, these results have been obtained with the standard theory of elasticity which implies to work with small deflections of the substrate surface. The limit $\gamma_{SL} \to 0$ for instance, that could be relevant for water deposited on a hydrogel (the two phases are very similar, which favours $\gamma_{SL}\approx 0$), is out of reach of this theory that seems to imply a value for $\theta_{SL}$ close to $90°$, that is here only an extrapolation.

**4-3 – Comparison with a more direct calculation of substrate slopes.**

Our derivation of eqs.(38) may seem rather complex, and one may suspect that a so simple result could perhaps be reached by an independent method. Inspired by the recent paper from Style and Dufresne [22], one can conjecture here that, near contact line the substrate should behave a bit like a liquid, satisfying both Neumann equilibria in the vertical and horizontal directions, while for consistency with the large scale behaviour, $\theta_0$ should coincide with the Young value $\theta_Y$, as we assumed in the previous section, too. This gives the set of three equations:

$$\begin{cases} \gamma \sin\theta_0 = \gamma_S \sin\theta_S + \gamma_{SL} \sin\theta_{SL} & (39-a) \\ \gamma \cos\theta_0 = \gamma_S \cos\theta_S - \gamma_{SL} \cos\theta_{SL} & (39-b) \\ \gamma \cos\theta_0 = \gamma_S - \gamma_{SL} & (39-c) \end{cases}$$

Here, as in the whole paper, we consider the limit $\gamma \ll \gamma_S$ and $\gamma \ll \gamma_{SL}$, which implies that the two substrate slope angles are small. In this limit, the two last equations give trivially:

$$\theta_{SL} \approx \sqrt{\frac{\gamma_S}{\gamma_{SL}}} \theta_S \qquad (40)$$

Using this equation in (39-a) give finally:

$$\begin{cases} \theta_S \approx \dfrac{\gamma \sin\theta_0}{\gamma_S + \sqrt{\gamma_S \gamma_{SL}}} & (41-a) \\ \theta_{SL} \approx \dfrac{\gamma \sin\theta_0}{\gamma_{SL} + \sqrt{\gamma_S \gamma_{SL}}} & (41-b) \end{cases}$$

that is very similar to (38) with the right symmetry on the indices S and SL, but is also clearly different... In particular the ratio $\theta_{SL}/\theta_S$ scales as $\sqrt{\gamma_S/\gamma_{SL}}$ instead of $\gamma_S/\gamma_{SL}$. It is however worth now to remind that we had also in the previous sections a second small parameter, that is the substrate surface tension mismatch: $\delta\gamma/\gamma_S=(\gamma_S-\gamma_{SL})/\gamma_S$. If we develop now eqs(41) upon this second small parameter, we get:

$$\begin{cases} \theta_S = \dfrac{\gamma \sin\theta_0}{2\gamma_S}\left(1 + \dfrac{1}{4}\dfrac{\delta\gamma}{\gamma_S}\right) & (42-a) \\ \theta_{SL} = \dfrac{\gamma \sin\theta_0}{2\gamma_{SL}}\left(1 - \dfrac{1}{4}\dfrac{\delta\gamma}{\gamma_{SL}}\right) & (42-b) \end{cases}$$

that satisfies again the symmetry exchange between the indices S and SL. As appears in these equations, our result was not so bad, but was differing from the exact value by a second order term, that is the cross-product of the two small parameters used in this paper. Our calculation is thus correct, but not enough accurate to predict delicate things like the ratio of the two slopes. Indeed, (38) would leed to $\theta_{SL}/\theta_S \approx 1 - \delta\gamma/\gamma_S$ instead of $\theta_{SL}/\theta_S \approx 1 - (1/2)\delta\gamma/\gamma_S$.

The reason of this discrepancy is presumably as follows. As reminded more above, the standard theory of elasticity is a first order linear theory, which prevents us, for consistency between bulk and surface equations, to develop (39-b) upon $\theta_S$ and $\theta_{SL}$ at more than order one. At this order, Neumann equation in the direction parallel to the substrate can not be distinguished from Young equation. As the exact calculation uses the difference of these two equations, our modelling can not recover exactly the full result. A complete treatment would require here to use a more ambitious modelling in terms of finite deformation of an hyperelastic medium, able to address larger values of the substrate slope, in a way similar to what has been done by Mora et al. [33] for Biot surface instabilities of a compressed medium. Though being certainly the most natural next step of the present approach, such a complex calculation is out of the scope of the present paper. Despite this problem, our result was not so far from the good one, with the right dependence upon the mean surface tension of the substrate and the good tendency for the dependence upon substrate surface tension mismatch. Our modelling is also able to give simple expressions for the typical displacements of the substrate in the direction normal to the surface that are presumably less sensitive to the difficulty to handle larger slopes of the substrate surface.

**5 – Wetting hysteresis in the sense of Extrand and Kumagai.**
The notion of local slope of the substrate, as addressed in the previous sections, play a central role in a model of static hysteresis developed long ago by Extrand and Kumagai [8] for soft substrates, with an application to elastomers in mind. These authors proposed a very simple model based on the implicit idea that the liquid does not always move coherently with the substrate distortion, and when it advances should "deal with" a condition of motion over a inclined surface, on the relevant side of the ridge (dry for advancing front, wet for receding fronts). Similar ideas have been recently used by Kajiya et al. [24] to interpret stick-slip motions of the contact line observed on gels having a less simple rheology. We first remind the ideas underlying their argument (subsection 5-1). We then use the results that we have obtained in the previous sections to see how this argument should be modified to include the existence of the substrate surface tension (subsection 5-2).

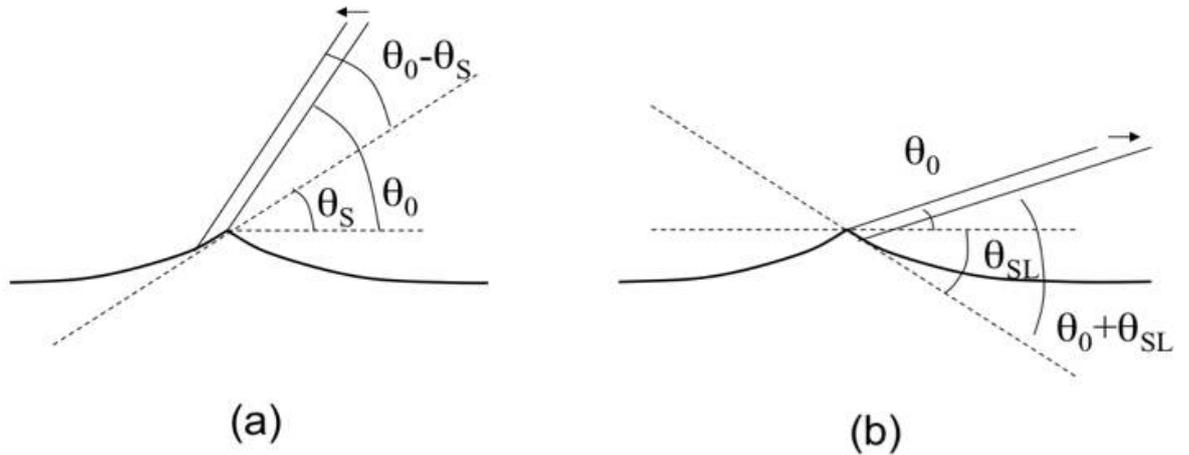

**Fig. 5:** *(a) In Extrand and Kumagai point of view, the liquid can advance ahead of the contact line (before relaxation of the substrate to a new shape), when the effective contact angle $\theta_0-\theta_S$ defined with respect to the inclined dry substrate becomes larger than a reference critical value $\theta_a^{(0)}$. (b) A similar argument can be built for receding of the liquid, the condition becoming now $\theta_0+\theta_{SL}<\theta_r^{(0)}$.*

The goal is not here to solve in details the problem addressed by Extrand and Kumagai (as we shall see there are difficulties in this description), but just to see the implications of our results for this specific kind of modelling.

**5-1 – Reminding Extrand and Kumagai argument.**

Though not initially provided with so much details, the argument proposed by Extrand and Kumagai [8] can be exposed as follows. Let us consider a drop that one progressively inflates on a deformable solid. On a purely elastic solid, a ridge should form and should accompany the motion of the contact line, as the typical velocity of surface waves is most often much larger than the advancing contact line velocity . If this velocity is small enough, dynamics effects on the value of the contact angle can be omitted, the contact angle being thus locked to the Young value while drop inflation takes place. This picture is true for a contact line having no pinning on the free surface of the substrate. If, now, on the contrary, one consider a *rigid* substrate that presents some pinning of the contact line, another state can be imagined in which the contact line does not move (pinned contact line) while the apparent contact angle does increase, until it reaches a critical value called $\theta_a^0$, before to advance again quasistatically with $\theta_0=\theta_a^0$ If one now mixes the two properties (intrinsic hysteresis of the substrate surface and deformability of the substrate), the advancing condition should presumably be defined with respect to the "dry" side of the ridge, which is inclined of an angle $\theta_S$. In this case a "catastrophe" can happen when the contact angle reaches $\theta_0=\theta_a^0+\theta_S$, as the liquid is in a unstable situation on the "dry" side of the ridge (see fig. 5-a). If it advances just a bit on the ridge, and if the ridge does not reorganize in time, the apparent value of $\theta_S$ will decrease, which drives again more the system above the depining condition. It is why, they propose a new depining condition for deformable substrate, that reads for the advancing case:

$$\theta_a = \theta_a^{(0)} + \theta_S \tag{43-a}$$

In the receding case (fig. 5-b), a similar argument can be developed on the "wet" part of the ridge, that leads to:

$$\theta_a = \theta_r^{(0)} - \theta_{SL} \tag{43-b}$$

Note that this argument is only a criterion for onset of an out of equilibrium situation. We do not know what will be the evolution of the system after depining, that should mix both the motion of the contact line and the reorganisation of the rim that will accompany its motion. This would merit a specific modelling, that was in fact tried by Carré and Shanahan [6], and byLong et al. [9] by adding some plasticity to the pure elastic modelling without substrate surface tension. As we said above, as the ridge can accompany the contact line in a pure elastic solid, some plasticity is essential to explain why the ridge could remain static while the liquid explores its sides. Interestingly, calculations of Carré and Shanahan, when extrapolated to a zero velocity of contact line predicts a residual static hysteresis, that has never been compared in detail with that of Extrand and Kumagai. The influence of yield and plasticity in Extrand and Kumagai description is certainly the first missing link to this theory.

Now, Extrand and Kumagai who tried to use Shanahan description of contact line on a soft solid were also puzzled by the fact that the solution is diverging near contact line which implies infinite values for $\theta_S$ and $\theta_{SL}$. They solved this empirically by observing that, on their data all seemed to happen as if there exists some cut-off of the divergence, so that (43) finally reduced to:

$$\theta_a = \theta_a^{(0)} + \frac{1}{2\pi}\frac{\gamma \sin\theta_a}{\mu x_0} \tag{44-a}$$

$$\theta_r = \theta_r^{(0)} - \frac{1}{2\pi}\frac{\gamma \sin\theta_r}{\mu x_0} \tag{44-b}$$

in which $x_0$ is a unknown cut-off at small scale of the divergence of the profile, found to be in the micron range. This description worked quite well in the elastomer case explored by Extrand and Kumagai, and is very attractive because of its simplicity, but the values found for $x_0$ were a bit large for a microscopic cut-off (micron size) and were never really interpreted. Also, as we have explained, surface tension of the substrate is completely omitted in this approach, though for the very soft materials often encountered in numerous applications, this parameter should play an essential role.

**5-2 – Implications of the present work.**

We thus reconsider Extrand and Kumagai approach in light of the modifications that we have added above to Shanahan initial description of elastowetting, for simplicity in the limit $R>>l_S$. Using eqs. (38), the advancing and receding angles, defined with respect to the horizontal should now be solutions of the transcendental equations:

$$\theta_a = \theta_a^{(0)} + \frac{\gamma \sin\theta_a}{\gamma_S + \sqrt{\gamma_S \gamma_{SL}}} \approx \theta_a^{(0)} + \frac{\gamma}{2\gamma_S}\sin\theta_a(1+\frac{1}{4}\frac{\delta\gamma}{\gamma_S}) \tag{45-a}$$

$$\theta_r = \theta_r^{(0)} - \frac{\gamma \sin\theta_r}{\gamma_{SL} + \sqrt{\gamma_S \gamma_{SL}}} \approx \theta_a^{(0)} - \frac{\gamma}{2\gamma_{SL}}\sin\theta_r(1-\frac{1}{4}\frac{\delta\gamma}{\gamma_{SL}}) \tag{45-b}$$

Note here that, if we forget the small parameters $\delta\gamma/\gamma_S$ and $\delta\gamma/\gamma_{SL}$ these equations can also been reached very easily by replacing the scale $x_0$ with the elastocapillary lengths $l_S=(1/\pi)(\gamma_S/\mu)$ and $l_{SL}=(1/\pi)(\gamma_{SL}/\mu)$, the $1/|x|$ profiles for the substrate slope being replaced by profiles of the kind $1/(l_S+|x|)$. This would be consistent with the simple idea that the most natural cut-off is here the elastocapillary length itself, that has the right order of magnitude (typically, with $\gamma_S \approx 30$ mN/m and $E \approx 0.3$ MPa, one gets $\gamma_S/E \approx 0.1\ \mu m$ ). If we now restrict to small values of our two parameters and also of all the angles involved here, one gets finally the following expressions for advancing and receding contact angles:

$$\theta_a \approx \theta_a^{(0)}\left[1+\frac{\gamma}{2\gamma_S}(1+\frac{1}{4}\frac{\delta\gamma}{\gamma_S})\right] \tag{46-a}$$

$$\theta_r \approx \theta_r^{(0)}\left[1-\frac{\gamma}{2\gamma_{SL}}(1-\frac{1}{4}\frac{\delta\gamma}{\gamma_{SL}})\right] \tag{46-b}$$

As one can see, this modelling leads to a rather strong hysteresis, the advancing contact angle being potentially much larger than the receding angle. A big surprise, however, with (46) is that it predicts a hysteresis that should only depend on the ratios of surface tension, and not upon plasticity and elasticity of the substrate, though obviously the coexistence of both effects are necessary to observe it. Also, the law (46) is unable to recover the absence of hysteresis that one would expect for a rigid material. This surprising result is not necessary unphysical as we are not in fact really dealing with a simple elastic substrate. We have invoked since the beginning some plasticity, governed by some specific time scale, to have a ridge that could not have time enough to adjust its shape while the liquid is exploring its sides. It is thus obvious that the limit reached here must differ from what would give a pure elastic medium, for which this kind of wetting hysteresis should even not exist.

Finally, comparisons of this approach with the data obtained by Extrand and Kumagai on elastomers [8] are difficult. First, eqs (46) can not be used directly because the contact angles are often not in the small angle limit. One should in this case use rather eqs. (45), that have a structure very similar to those used by Extrand and Kumagai (eqs. 41). However, we do not know precisely the dependence of $\gamma_S$ and $\gamma_{SL}$ upon the degree of reticulation. Now, if we suppose that, being governed basically by the same microscopic interactions, these two quantities have dependences similar to that of $\mu$, the quality of the description should be the same for eqs (41) and (45), the second ones avoiding to introduce some unknown length scale $x_0$. With this respect, we have perhaps here improved Extrand and Kumagai modelling, but further explorations of wetting hysteresis on soft substrates would be welcome, with combined investigation of the distortion profiles, to measure explicitly the scales $l_S=(1/\pi)\gamma_S/\mu$ and $l_{SL}=(1/\pi)\gamma_{SL}/\mu$ as well as the angles $\theta_S$ and $\theta_{SL}$.

**6-Conclusion**

In summary, this paper has reconsidered the problem of elastic distortions induced by a contact line on a soft solid, and performed its extension to two parallel contact lines, while including the effect of the substrate surface tension in the incompressibility limit. Several results have been obtained:

- We have shown that, with a reasonably good accuracy, the surface of the substrate can be described with logarithmic profiles, except that the divergences at respectively large and small scales are cut respectively by the distance between the two contact lines (or any other macroscopic limit, more generally) and the elastocapillary length. At small length scales, the divergence of the elastic solution is replaced by an equilibrium of the three surface tensions that meet at the contact line, just as for liquids, in agreement with what found Jerison et al. [18].

- We have calculated explicitly the distortion of the substrate for one or two contact lines, giving formula for the height of the ridge, slope of its two sides (wetted and non wetted), deepness of the depletion induced by Laplace pressure below the drop, etc… Though most of the paper deals with the "symmetric" case in which the substrate surface tension is the same for both the dry and wetted regions, we have been able to propose an approximate extension to the "asymmetrical" case, and even to calculate the limit slope of each substrate part (dry and wetted) near contact line.

- Finally, we have reconsidered a hysteresis criterion, invoked long ago by Extrand and Kumagai [8], and explored the implications of our calculations on this one. Our approach is equivalent to replace the unknown length scale identified by these authors by the elastocapillary length [25]. A surprising result is here that the final value we found seem to depend only upon the ratios of surface tensions involved, though elasticity as well as plasticity are essential to generate such an hysteresis linked to the geometry of the distorted solid substrate.

It would be now interesting to measure extensively the wetting hysteresis presented by various soft substrates (polymers, gels, elastomers…) and to try to correlate the results with their surface tension, as well as with their elastic and plastic properties. Surface tensions of the substrate are not always easy to measure, especially the mismatch between wet and dry values, though forcing of waves at the substrate surface by acoustics or optics could offer some means to measure *in situ* both bulk rheology and surface properties [34]. In another direction, the theory developed in this paper is interesting because of the reduction of elasticity to a Stokes framework. After an appropriate combination with viscous degrees of freedom, remaining in a similar framework, this could open simple ways to model viscoelastic substrate, and thus to address the complex behaviours reported in recent stick-slip wetting experiments [13-14,24]. The next step would then consist in including the possible flow of the liquid inside the substrate (permeation, swelling) which also leeds to specific wetting effects, as discussed recently by Kajiya et al. [19], or the effect of a surfactant at the free surface which has also motivated several experimental studies on gel substrates [10-12, 16] with puzzling self-organisation phenomena, still not well understood. It would also be interesting to see how the limit explored in the present problem could match to what happen on very thin elastic layers [9,18], and also to the case of a "free" sheet, with both bending and buckling instead of shear or elongation effects [35-36]. Finally, experiments with systems having only elastic properties even at very large strain, as the telomers developed recently [37], would be welcome, as well as more advanced modelling on the basis of viscoelastic Maxwell solids [38], or hyperelastic solids when strong deformations are involved [33].

**Acknowledgments:** the approximate solution (6) has been imagined during discussions with M. Banaha and A. Daerr, and is yet qualitatively explained in ref. [17]. I am also indebted to discussions and exchanges with J. Dervaux, J. Eggers, T. Kajiya, F. Lequeux, A. Marchand, J. Moukhtar, and C. T. Pham.